\pdfoutput=1
\documentclass[preprint,prd,preprintnumbers,amsmath,amssymb,superscriptaddress,nofootinbib]{revtex4}

\usepackage{epsfig,psfrag,graphics,verbatim}
\usepackage{graphicx}% Include figure files
\usepackage{dcolumn}% Align table columns on decimal point
\usepackage{bm}% bold math
\usepackage{slashed}
\usepackage{color}

\begin{document}
\hfill{{\small TUM-HEP 836/12}}

\title{Dark matter constraints from box-shaped gamma-ray features}

\author{Alejandro Ibarra}
\affiliation{Physik-Department T30d, Technische Universit\"at M\"unchen, James-Franck-Stra\ss{}e, 85748 Garching, Germany}

\author{Sergio L\'opez Gehler}
\affiliation{Physik-Department T30d, Technische Universit\"at M\"unchen, James-Franck-Stra\ss{}e, 85748 Garching, Germany}

\author{Miguel Pato}
\affiliation{Physik-Department T30d, Technische Universit\"at M\"unchen, James-Franck-Stra\ss{}e, 85748 Garching, Germany}

\date{\today}% It is always \today, today,

\begin{abstract}
The observation of a sharp spectral feature in the gamma-ray sky would be one of the cleanest ways to identify dark matter and pinpoint its properties. Over the years a lot of attention has been paid to two specific features, namely gamma-ray lines and internal bremsstrahlung. Here, we explore a third class of spectral signatures, box-shaped gamma-ray spectra, that naturally arise in dark matter cascade annihilations or decays into intermediate particles that in turn decay into photons. Using Fermi-LAT data, we derive constraints on the dark matter parameter space for both annihilating and decaying dark matter, and show explicitly that our limits are competitive to strategies employing standard spectral features. More importantly, we find robust limits even in the case of non-degenerate dark matter and intermediate particle masses. This result is particularly relevant in constraining dark matter frameworks with gamma-ray data. We conclude by illustrating the power of box-shaped gamma-ray constraints on concrete particle physics scenarios.
\end{abstract}

%\pacs{}% PACS, the Physics and Astronomy
%                             % Classification Scheme.
\maketitle

\section{Introduction}
\par The quest for dark matter (DM) is entering a crucial period as collider, direct and indirect searches close in on the most popular frameworks for weakly interacting massive particles (WIMPs). No clear, unambiguous signal has been pinpointed as yet and correspondingly stringent constraints on the properties of dark matter have been derived. Indirect searches, in particular, offer a multitude of channels and potential targets as discussed at length in the literature (see for instance \cite{BertoneReview,BergstromReview,BertoneBook,BergstromMulti} for reviews). However, indirect signals are, on the one hand, prone to different sorts of uncertainties (including cosmic-ray propagation \cite{SM98,Donato2001,Delahaye07,dragon2,Putze1,Trotta:2010mx} and the dark matter distribution in our Galaxy \cite{Navarro:2003ew,Gao:2007gh,Diemand:2008in,Navarro:2008kc}) and, on the other hand, lack strong discrimination power against astrophysical sources \cite{grasso,elecAMS02}. This makes any claim of identification of dark matter through indirect channels -- as well as the reconstruction of the underlying properties -- extremely problematic. In this context, gamma-ray spectral features assume today a particularly relevant role since there is no known astrophysical source capable of mimicking such signature. Moreover, the LAT instrument \cite{Atwood:2009ez,fermilatsite} on board the Fermi satellite has dramatically improved our mapping of the gamma-ray sky and offers unprecedented sensitivity in the search for spectral features. Therefore, gamma-ray signatures rest as perhaps the best indirect strategy to identify dark matter in an unambiguous fashion. Signatures such as gamma-ray lines at (a fraction of) the dark matter particle mass $m_{DM}$ and the hard spectra provided by internal bremsstrahlung just below $m_{DM}$ have been thoroughly investigated \cite{BergstromSnellman,BergstromUllioBuckley,Gustafsson,Bertone:2009cb,VertongenWeniger,Bringmannetal} and shown to place strong limits on particle physics models (for a tentative gamma-ray line, see the recent works \cite{Weniger:2012tx,gamma130Raidal}). Here, we focus on a different spectral feature that is somewhat unexplored in the existing literature. If dark matter annihilates or decays into an intermediate particle $\phi$ that in turn decays appreciably into photons, the resulting photon spectrum resembles a ``box'' whose centre and width are entirely fixed by $m_{DM}$ and $m_{\phi}$. The aim of the present paper is to study the relevance of these box-shaped features for dark matter searches and to show that they can be at least as constraining as the widely-studied gamma-ray lines. We start by discussing the phenomenology of dark matter cascade processes and the corresponding photon fluxes in Section \ref{secGamma}. In Section \ref{secResults}, we derive our constraints on the dark matter parameter space using Fermi-LAT data and briefly address the detectability of box-shaped gamma-ray features. Finally, we present concrete particle physics models for cascade annihilations into photons in Section \ref{secModels}, before concluding in Section \ref{secConcl}.

\section{Gamma-ray spectral features from cascade annihilations or decays}\label{secGamma}

\par Dark matter cascade annihilations into light degrees of freedom have been explored before \cite{Pospelov:2007mp,Cholis:2008vb,Nomura,Cholis:2008wq,Bergstrom:2008ag,Nomura2} as a means of yielding sizable fluxes of electrons and positrons, diffuse gamma-rays, synchrotron radiation and neutrinos. We focus instead on the possibility of producing distinctive gamma-ray spectral features through 1-step cascade processes. Suppose that dark matter self-annihilates into a pair of scalars $\phi$ that in turn decay into a pair of photons\footnote{The case of dark matter self-annihilation into one scalar and another particle and/or the decay of the scalar into one photon and another particle is also feasible, but leads to very similar phenomenology as presented here and is therefore omitted.} (for concrete particle physics realisations and branching ratios see Section \ref{secModels}). Each of the four photons emitted per annihilation has a monochromatic energy $E_{\gamma}'=m_{\phi}/2$ in the rest frame of the corresponding scalar $\phi$. In the lab frame -- where the dark matter particles are non-relativistic and the scalars have energy $E_{\phi}=m_{DM}$ -- the photon energy reads

\begin{equation}\label{ephoton}
E_\gamma = \frac{m_{\phi}^2}{2\,m_{DM}}\left(1-\cos\theta\sqrt{1-\frac{m_{\phi}^2}{m_{DM}^2}}\right)^{-1} \quad ,
\end{equation}
with $\theta$ the angle between the outgoing photon and the parent scalar in the lab frame. From this equation we read that the spectrum has sharp ends defined by the parameters $m_{DM}$ and $m_{\phi}$. The highest (lowest) energy corresponds to a photon emitted at an angle $\theta=0^{\circ}$ ($180^\circ$) with respect to the momentum of the parent scalar. Since the decaying particle is a scalar, the photon emission is isotropic. Hence the resulting spectrum is constant between the energy endpoints and takes a flat, box-shaped form:
\begin{equation}\label{box}
\frac{dN_\gamma}{dE_\gamma} = \frac{4}{\Delta E} \Theta(E-E_{-}) \Theta(E_{+}-E) \quad ,
\end{equation}
where $\Theta$ is the Heaviside function, $\Delta E=E_{+}-E_{-}=\sqrt{m_{DM}^2-m_\phi^2}$ is the box width and $E_{\pm}=(m_{DM}/2)\left(1\pm \sqrt{1-m_{\phi}^2/m_{DM}^2}\right)$ are the box edges. In the case of dark matter decays into a pair of scalars, the above expressions apply with the replacement $m_{DM}\to m_{DM}/2$.

\par Let us briefly discuss the impact of the primary spectrum \eqref{box} on gamma-ray searches for dark matter. First of all, notice that the centre of the ``box'' $E_c\equiv(E_{+}+E_{-})/2$ lies at $m_{DM}/2$, while the relative width is chiefly regulated by the degeneracy parameter $\Delta m/m_{DM}\equiv(m_{DM}-m_\phi)/m_{DM} \in [0,1]$ since $\Delta E/E_c=2\sqrt{2\Delta m/m_{DM}-\Delta m^2/m_{DM}^2}$. Hence, $m_{DM}$ and $\Delta m/m_{DM}$ fix the shape of the boxy spectrum, and accordingly we shall use both as our main phenomenological parameters. Secondly, in the limit $\Delta m/m_{DM} \to 0$ (or, effectively, when $\Delta E/E_c$ falls bellow the experimental resolution) the spectrum $dN_\gamma/dE_\gamma$ reduces to a line. But, unlike the well-known $\gamma\gamma$ and $\gamma Z$ lines, in this case there are four photons emitted per annihilation and each carries half of the dark matter particle rest-mass energy, or one-fourth for decaying dark matter models. Clearly, as we consider less and less degenerate configurations, i.e.~as $\Delta m/m_{DM}\to 1$, the gamma-ray box gets wider in energy and dimmer in amplitude which might lead one to conclude that non-degenerate cases are hardly observable. However, as we shall see, even for $\Delta m/m_{DM}=1$ the spectral plateau sits at non-negligible amplitudes and its feebleness is partially compensated by the extension to high energies. In fact, given that all known gamma-ray backgrounds (and also some spectral features such as internal bremsstrahlung) fall rapidly with energy, a flat spectrum at high energies -- albeit dim -- has good detectability prospects and strong constraining power. This is the backbone of our approach and shall be explicitly illustrated in Section \ref{secResults}.

\par Now, the flux of photons at the Earth induced by the annihilation spectrum \eqref{box} in the case of Majorana dark matter particles is given by \cite{BertoneReview,BergstromReview,VertongenWeniger}
\begin{equation}\label{unconvol}
\tilde{\phi}_\gamma(E_\gamma)\equiv \frac{d^4 N_\gamma}{dE_\gamma dS d\Omega dt} = \frac{\langle \sigma v \rangle}{8\pi m_{DM}^2} \, \frac{dN_{\gamma}}{dE_{\gamma}} \, \frac{1}{\Delta \Omega} \int_{\Delta \Omega}{d\Omega \, J_{ann}} \quad ,
\end{equation}
where $\langle \sigma v \rangle$ is the thermally averaged annihilation cross-section into $\phi$ pairs, $\Delta \Omega$ is the observed field of view and $J_{ann}=\int_{l.o.s.}{ds \, \rho_{DM}^2}$ is the integral of the squared dark matter density $\rho_{DM}$ along the line of sight. For models where the dark matter particle is not Majorana, equation \eqref{unconvol} is to be multiplied by a factor 1/2. The photon flux induced by decaying dark matter is obtained by replacing $\langle \sigma v \rangle/8\pi m_{DM}^2 \to \Gamma/4\pi m_{DM}$ ($\Gamma$ is the decay rate) and $J_{ann} \to J_{dec}=\int_{l.o.s.}{ds \, \rho_{DM}}$ in equation \eqref{unconvol}, as well as $m_{DM}\to m_{DM}/2$ in equations \eqref{ephoton} and \eqref{box}. In any spectral analysis it is essential to convolute the expected flux $\tilde{\phi}_\gamma$ with the experimental energy resolution. Assuming a gaussian detector response characterised by a standard deviation $\sigma(E)$, the convoluted photon flux reads
\begin{equation}\label{convol}
\phi_\gamma (E_\gamma) = \int{dE_\gamma' \,\, \tilde{\phi}_\gamma(E_\gamma') \frac{1}{\sqrt{2\pi} \sigma(E_\gamma')} \exp\left( -\frac{(E_\gamma'-E_\gamma)^2}{2\sigma^2(E_\gamma')} \right)  } \quad .
\end{equation}
This procedure smears all sharp edges and spikes eventually present in $\tilde{\phi}_\gamma$ and ultimately limits the search for spectral features.

\begin{figure}[htp]
\centering
\hspace{-1cm}
\includegraphics[width=0.52\textwidth,height=0.43\textwidth]{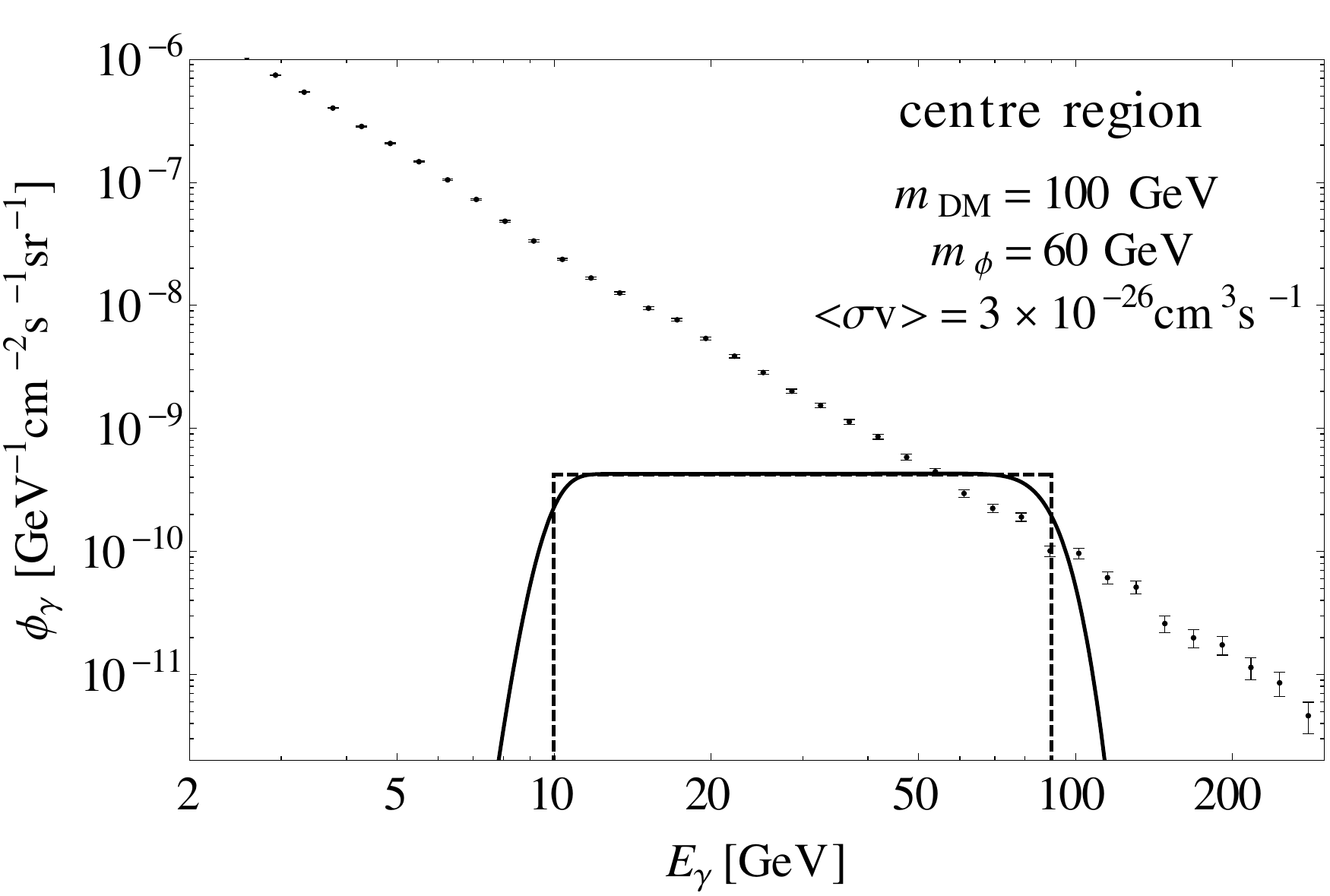}
\includegraphics[width=0.52\textwidth,height=0.43\textwidth]{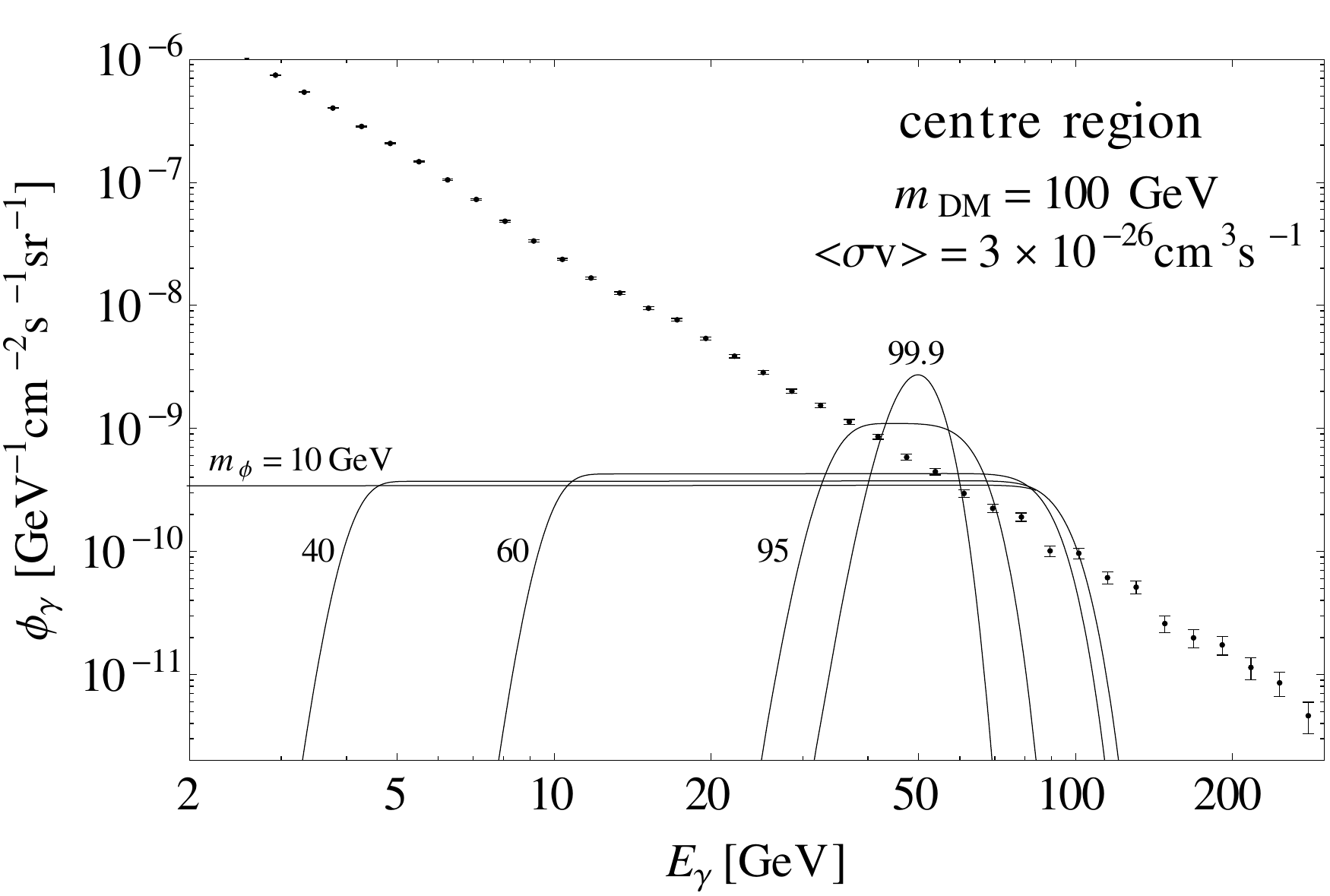}
\caption{The spectrum of box-shaped gamma-ray features. The left panel displays the unconvoluted (dashed) and convoluted (solid) box spectra for $m_{DM}=100$ GeV, $m_{\phi}=60$ GeV and $\langle \sigma v \rangle = 3\times 10^{-26} \textrm{ cm}^3/\textrm{s}$ on top of the 2-yr Fermi-LAT data (borrowed from \cite{VertongenWeniger}) for the galactic centre region. The right frame shows the convoluted box spectra for $m_{DM}= 100$ GeV, $\langle \sigma v \rangle = 3\times 10^{-26} \textrm{ cm}^3/\textrm{s}$ and several values of $m_{\phi}$.
}
\label{fig:fluxes}
\end{figure}

\par We are interested in exploring box-shaped gamma-ray features in the energy range relevant for typical WIMPs, i.e.~from a few GeV to a few TeV, so we shall focus on Fermi-LAT performance and data all through the manuscript. The energy resolution of the LAT instrument is assumed gaussian and parameterised according to \cite{Rando:2009yq,fermilatsite2}, giving $\sigma(E)/E=8 \, (12)\%$ at $E= 1 \, (200)$ GeV. Considering a more realistic non-gaussian LAT energy response would lead to a slightly different flux drop at the edges of the box spectrum; that would be relevant for the identification of a spectral feature but is of reduced importance in deriving flux constraints as we pursue in the present article. We consider as our main regions of interest the galactic centre and halo regions as defined in \cite{VertongenWeniger} (cf.~Table 1 therein). The former features $\Delta \Omega=1.30$ sr, $\int_{\Delta \Omega}{d\Omega \, J_{ann}}=9.2\times 10^{22}$ GeV$^2$ cm$^{-5}$ sr and $\int_{\Delta \Omega}{d\Omega \, J_{dec}}=6.9\times10^{22}$ GeV cm$^{-2}$ sr, while the latter presents $\Delta \Omega=10.4$ sr, $\int_{\Delta \Omega}{d\Omega \, J_{ann}}=8.3\times 10^{22}$ GeV$^2$ cm$^{-5}$ sr and $\int_{\Delta \Omega}{d\Omega \, J_{dec}}=2.2\times10^{23}$ GeV cm$^{-2}$ sr, assuming a Navarro-Frenk-White (NFW) profile normalised to a local dark matter density of $0.4$ GeV/cm$^3$. Following the findings of \cite{VertongenWeniger}, we shall focus on the centre (halo) region to derive constraints on annihilating (decaying) dark matter. For the centre region, figure \ref{fig:fluxes} (left) shows the unconvoluted and convoluted box spectra taking $m_{DM}=100$ GeV, $m_{\phi}=60$ GeV (or $\Delta m/m_{DM}=0.4$) and $\langle \sigma v \rangle = 3\times 10^{-26} \textrm{ cm}^3/\textrm{s}$, as well as the 2-yr Fermi-LAT data borrowed from the analysis in \cite{VertongenWeniger} and distributed into 20 energy bins per decade. We have checked a posteriori that our results do not vary significantly with the particular choice of the binning. The reader is referred to \cite{VertongenWeniger} for further details on the Fermi-LAT data used here. Figure \ref{fig:fluxes} (right) illustrates instead the effect of varying the mass degeneracy parameter $\Delta m/m_{DM}$. The plots highlight the key phenomenological features of the dark matter models under scrutiny. As discussed above, in the limit of vanishing $\Delta m$ the box spectrum squeezes to a gamma-ray line whose amplitude and width are fixed by the energy resolution $\sigma(E)$. Furthermore, the right panel of figure \ref{fig:fluxes} clearly shows the relative importance of a flat spectrum extending to high energies. Notice as well that the convoluted box is not exactly symmetric about its centre because of the energy-dependent $\sigma(E)/E$. Since the gamma-ray flux measured by Fermi-LAT is softer than $E_\gamma^{-2}$, we can already anticipate that the most constraining energy bin for a given box spectrum is usually the one at which $E_\gamma^2 \phi_\gamma$ peaks.

\par Before proceeding with the derivation of Fermi-LAT constraints, a few comments are in order. The shape of the dark matter profile entering in the $J-$factors is not precisely known as of today despite the ground-breaking progress in numerical simulations over the last decade or so \cite{Navarro:2003ew,Gao:2007gh,Diemand:2008in,Navarro:2008kc}. More or less cuspy profiles towards the galactic centre result in significantly different $J-$factors, so it has become customary to use NFW, Einasto and isothermal profiles in order to bracket dark matter constraints. This sort of uncertainty, together with the local dark matter density that is uncertain to tens of percent \cite{CaldwellOstriker,Gates1995,UllioBuckley,BelliFornengo,CatenaUllio,Weber:2009pt,SaluccilocalDM,paperDMlocal}, translates directly into the normalisation of the fluxes in figure \ref{fig:fluxes}. Likewise, the particular choice of the observed target region affects sensitively the derived gamma-ray constraints (see for instance \cite{Pieri:2009je,Ibarra:2009nw,Weniger:2012tx}). All this problematic has been exhaustively addressed in the literature; what we wish to motivate and pursue here is the search for a different kind of gamma-ray feature that is at least as powerful as the most well-studied spectral signatures.

\section{Results}\label{secResults}

\par It is already evident from figure \ref{fig:fluxes} that thermal cross-sections for dark matter annihilations into intermediate scalars are in tension with the Fermi-LAT observations. We now turn to the derivation of precise constraints on the dark matter parameter space. A given dark matter model defined by $(m_{DM},\langle \sigma v\rangle,\Delta m/m_{DM})$ -- or $(m_{DM},\Gamma,\Delta m/m_{DM})$ -- is considered excluded if the photon signal $\phi_\gamma$ plus background $\phi_{\gamma,b}$ exceed the measurement more than 2$\sigma$ in at least one Fermi-LAT energy bin over the range $1-280$ GeV. In order to bracket the uncertainty regarding background modelling, we adopt three distinct approaches: \emph{(i) conservative}, in which we boldly assume no background $\phi_{\gamma,b}=0$, naturally leading to the weakest limits; \emph{(ii) intermediate}, where we fit Fermi-LAT datapoints below $20$ GeV to a power-law (scaled down to $\sim 80$\% to give room to a putative dark matter signal) and take that to describe the gamma-ray background $\phi_{\gamma,b}\propto E^{-\nu}$ ($\nu\simeq2.6$); and \emph{(iii) aggressive}, where we take the background to meet the central points of the measurements which is a slightly overoptimistic method of setting constraints on dark matter models. Notice that the intermediate prescription represents an ad hoc approach of illustrative character and should be refined for more extensive analyses. This procedure, although not as sophisticated as others \cite{VertongenWeniger,Bringmannetal,Weniger:2012tx}, is appropriate for our purposes and enables an overall assessment of gamma-ray constraints on boxy spectra.

\begin{figure}[htp]
\centering
\hspace{-1cm}
\includegraphics[width=0.52\textwidth,height=0.43\textwidth]{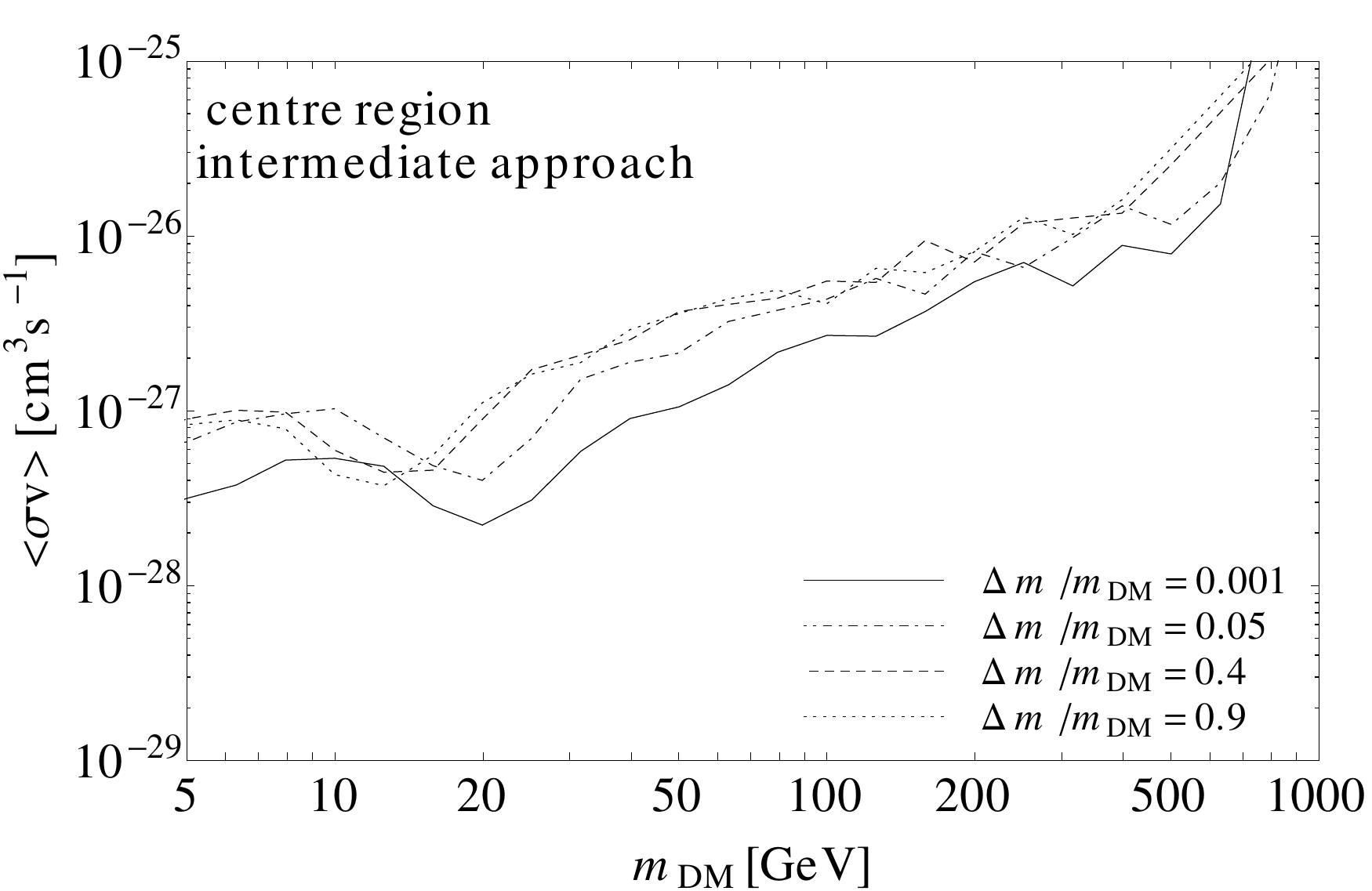}
\includegraphics[width=0.52\textwidth,height=0.43\textwidth]{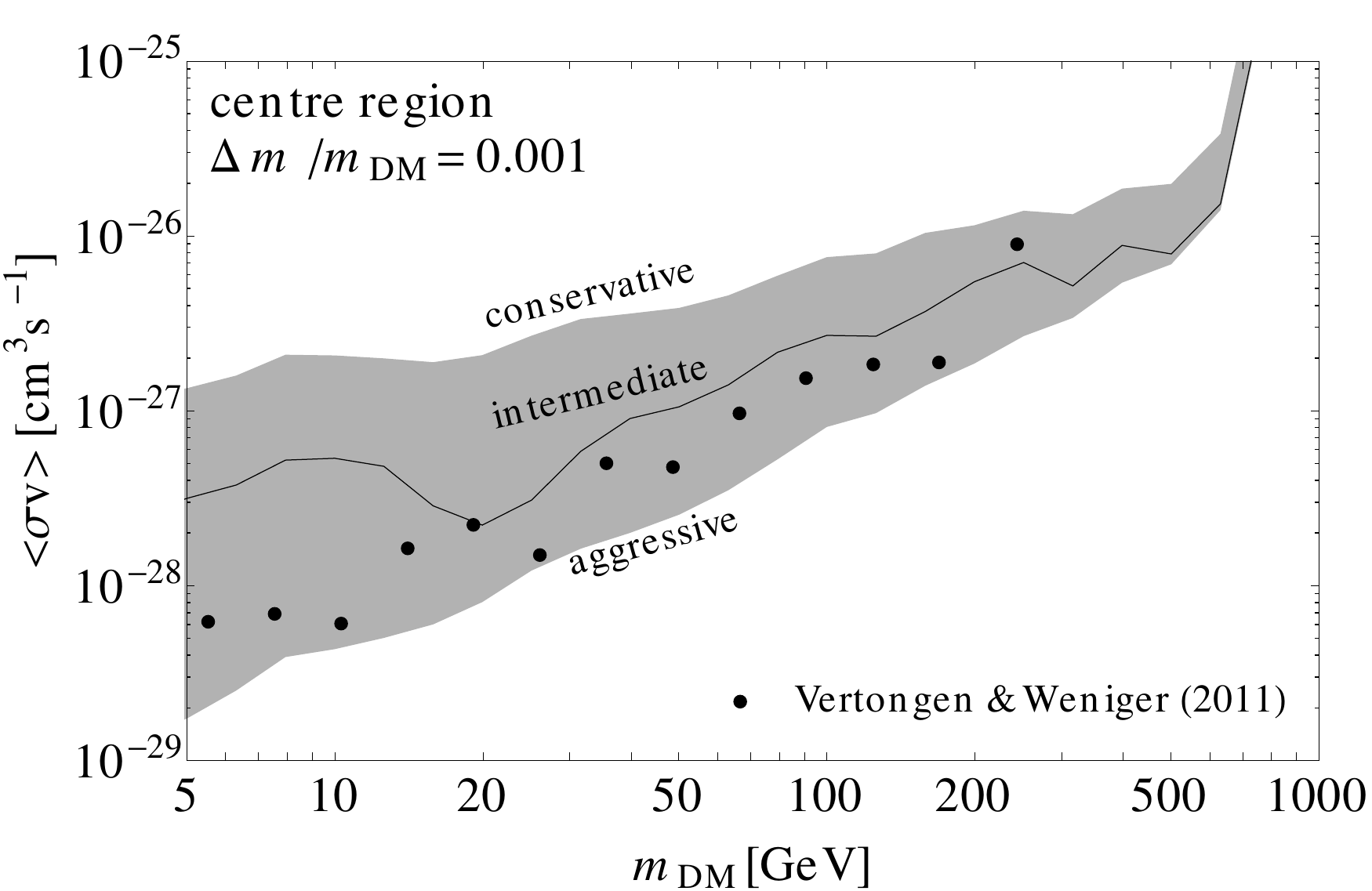}
\caption{
Gamma-ray constraints on cascade dark matter annihilations for the galactic centre region. In the left panel the lines show our constraints in the $\langle \sigma v \rangle$ vs $m_{DM}$ plane for several values of the degeneracy parameter $\Delta m/m_{DM}$ and using the intermediate approach. In the right panel we compare the constraints obtained with the different aproaches for $\Delta m/m_{DM}=0.001$. The constraints obtained in Ref.~\cite{VertongenWeniger} for $\gamma\gamma$ lines are represented by the filled circles.
}
\label{fig:constraints_ann}
\end{figure}

\par Figure \ref{fig:constraints_ann} (left) displays our constraints in the $\langle \sigma v \rangle$ vs $m_{DM}$ plane for different values of $\Delta m/m_{DM}$ and using our intermediate approach. The first overall impression is that the upper limits on the annihilation cross-section into a pair of scalars are rather strong, sitting up to two orders of magnitude below the usual thermal cross-section $\langle \sigma v \rangle=3\times 10^{-26}\textrm{ cm}^3/\textrm{s}$. Considering our aggressive approach the constraints exclude even lower cross-sections of about $10^{-29}\textrm{ cm}^3/\textrm{s}$ at the lowest dark matter masses, as it is shown on the right panel of figure \ref{fig:constraints_ann}. This appears to be a general feature of cascade models and, as we shall see in Section \ref{secModels}, sets stringent limits on sensible particle physics models. The dips seen at $m_{DM}\simeq 10-20$ GeV in the left panel of figure \ref{fig:constraints_ann} are artifacts of our intermediate approach and result from the energy bin at $E_{\gamma}=10$ GeV where the measured flux is closest to our assumed background power-law. This makes the intermediate constraints for the almost degenerate case particularly strong at $m_{DM}=20$ GeV for annihilating dark matter (since $E_{\gamma}=m_{DM}/2$) and $m_{DM}=40$ GeV for decaying dark matter (since $E_{\gamma}=m_{DM}/4$). The grey band in figure \ref{fig:constraints_ann} (right) brackets the constraints derived with our conservative, intermediate and aggressive approaches. The band encompasses almost two orders of magnitude at low masses and less than one at high masses. This is because in the aggressive case the constraints are set by the Fermi-LAT error bars, which are much smaller at lower energies. Notice as well that, in the almost degenerate case at $m_{DM}\gtrsim 600$ GeV, the low-energy end of the box surpasses the last bin of Fermi-LAT data, leading to extremely weak constraints and to the sudden rise in the high-mass region of both panels in figure \ref{fig:constraints_ann}.

\par Secondly, as expected, the strongest constraints are obtained for the almost degenerate case $\Delta m/m_{DM}=0.001$, in which the gamma-ray spectrum resembles a line. In this case, a comparison with the results of Ref.~\cite{VertongenWeniger} (shown by the filled circles in the right panel of figure \ref{fig:constraints_ann}) is allowed, but the translation of the results is not immediate. In fact, a direct annihilation into $\gamma\gamma$ has a monochromatic spectrum at $m_{DM}$ normalised to two photons per annihilation, while in the degenerate case of our model the line occurs at $m_{DM}/2$ and is normalised to four photons per annihilation.  In any case, as we can see on the right panel of figure \ref{fig:constraints_ann}, our aggressive approach gives similar results to the detailed likelihood method presented in \cite{VertongenWeniger}.

\par
\begin{figure}[htp]
\centering
\includegraphics[width=0.49\textwidth]{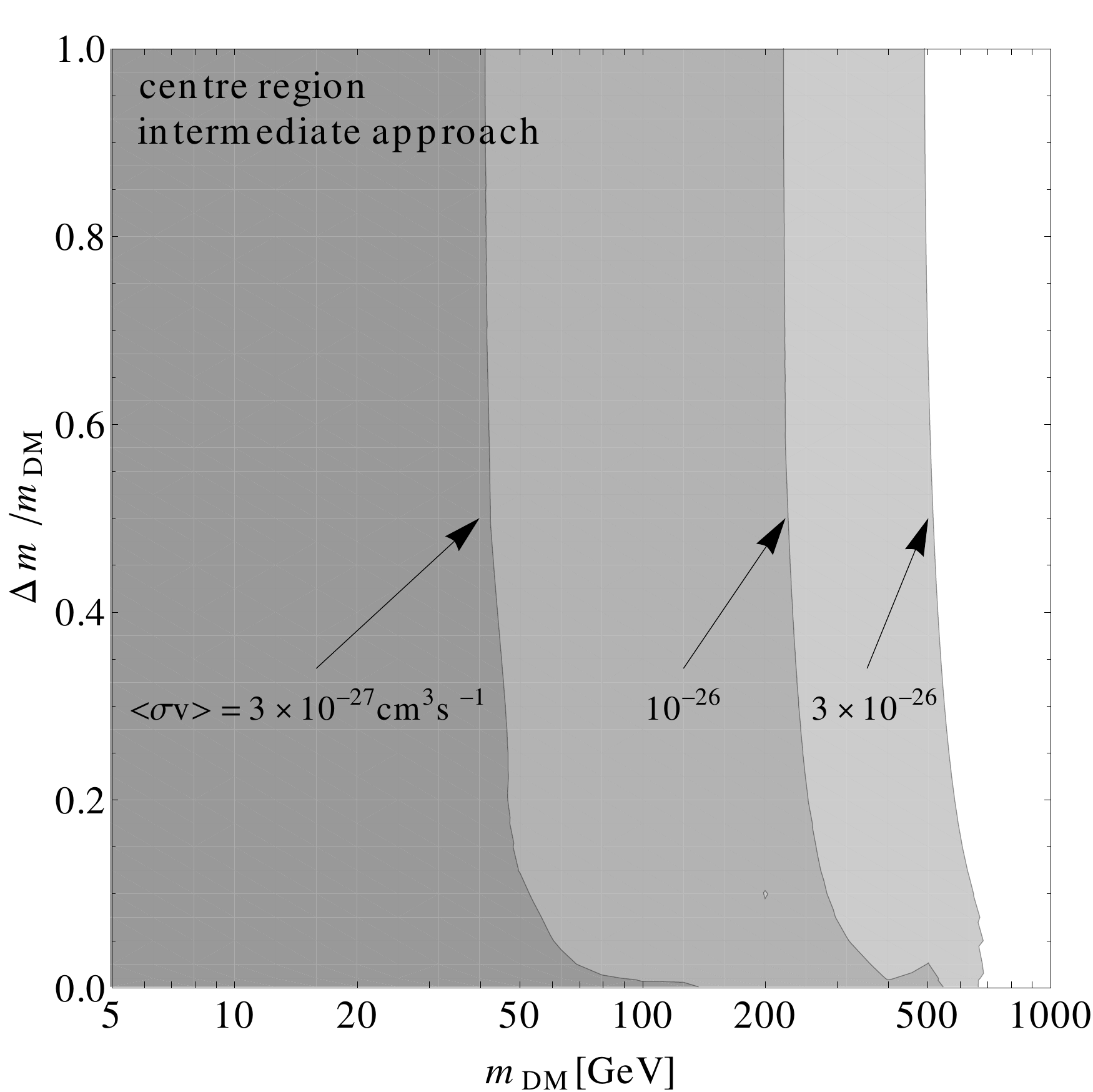}
\includegraphics[width=0.49\textwidth]{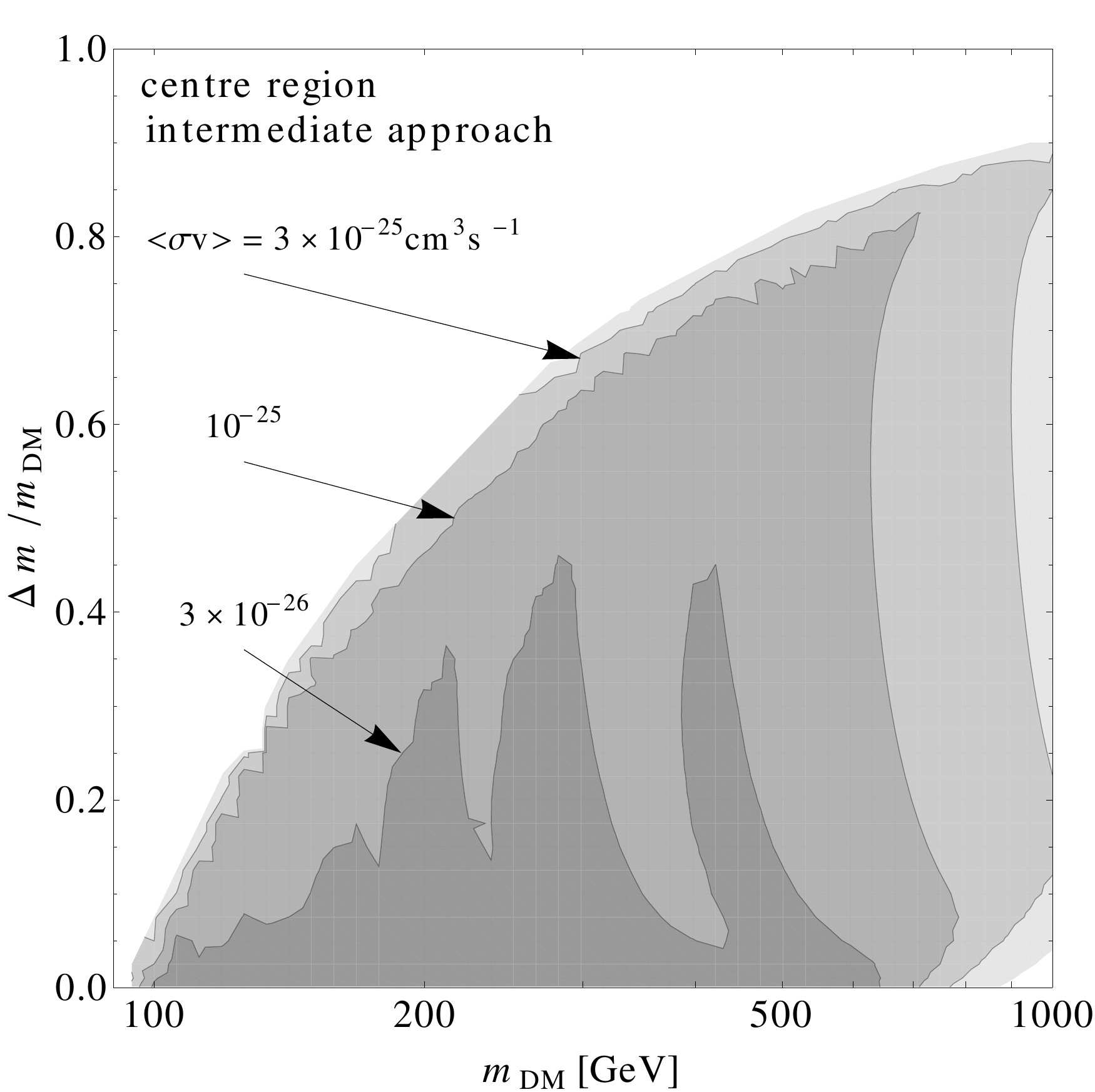}
\caption{
Annihilating dark matter constraints in the $\Delta m/m_{DM}$ vs $m_{DM}$ plane for different values of $\langle \sigma v \rangle$ using the intermediate approach. The shaded areas are excluded for the respective cross-sections. In the left (right) panel we show the case of intermediary particles decaying into $\gamma \gamma$ ($\gamma Z$).
}
\label{fig:contour}
\end{figure}

 Finally, let us notice that, as $\Delta m/m_{DM}\to 1$, the constraints in figure \ref{fig:constraints_ann} (left) weaken, but the weakening seems to ``saturate'' at the highest values of $\Delta m/m_{DM}$ leading to rather robust constraints even in those cases. The reason behind such behaviour has already been addressed in Section \ref{secGamma}: the decrease in the box amplitude as less and less degenerate cases are considered is compensated by the extension of the box edge $E_{+}$ to higher energies, where better sensitivities are achieved. This has the important consequence that, for virtually all relative degeneracies $\Delta m/m_{DM}\gtrsim 0.05$, one can exclude roughly the same range of dark matter masses for fixed $\langle \sigma v \rangle$, as more clearly seen in figure \ref{fig:contour} (left). There we show the excluded regions of the dark matter parameter space for different values of $\langle \sigma v \rangle$. For example, at $\langle \sigma v \rangle=3\times 10^{-26}\textrm{ }(3\times 10^{-27})\textrm{ cm}^3\textrm{/s}$ and $\Delta m/m_{DM}\gtrsim 0.05$ the dark matter mass range $m_{DM}\simeq 5-600$ ($5-50$) GeV is excluded using our intermediate approach. For completeness we show in figure \ref{fig:contour} (right) the excluded parameter space in the case of decays $\phi\to \gamma Z$, where the upper left white region is forbidden for kinematical reasons. In practice, even for particle physics realisations featuring highly non-degenerate configurations, it is possible to derive meaningful, stringent constraints.

\par
\begin{figure}[htp]
\centering
\includegraphics[width=0.52\textwidth,height=0.43\textwidth]{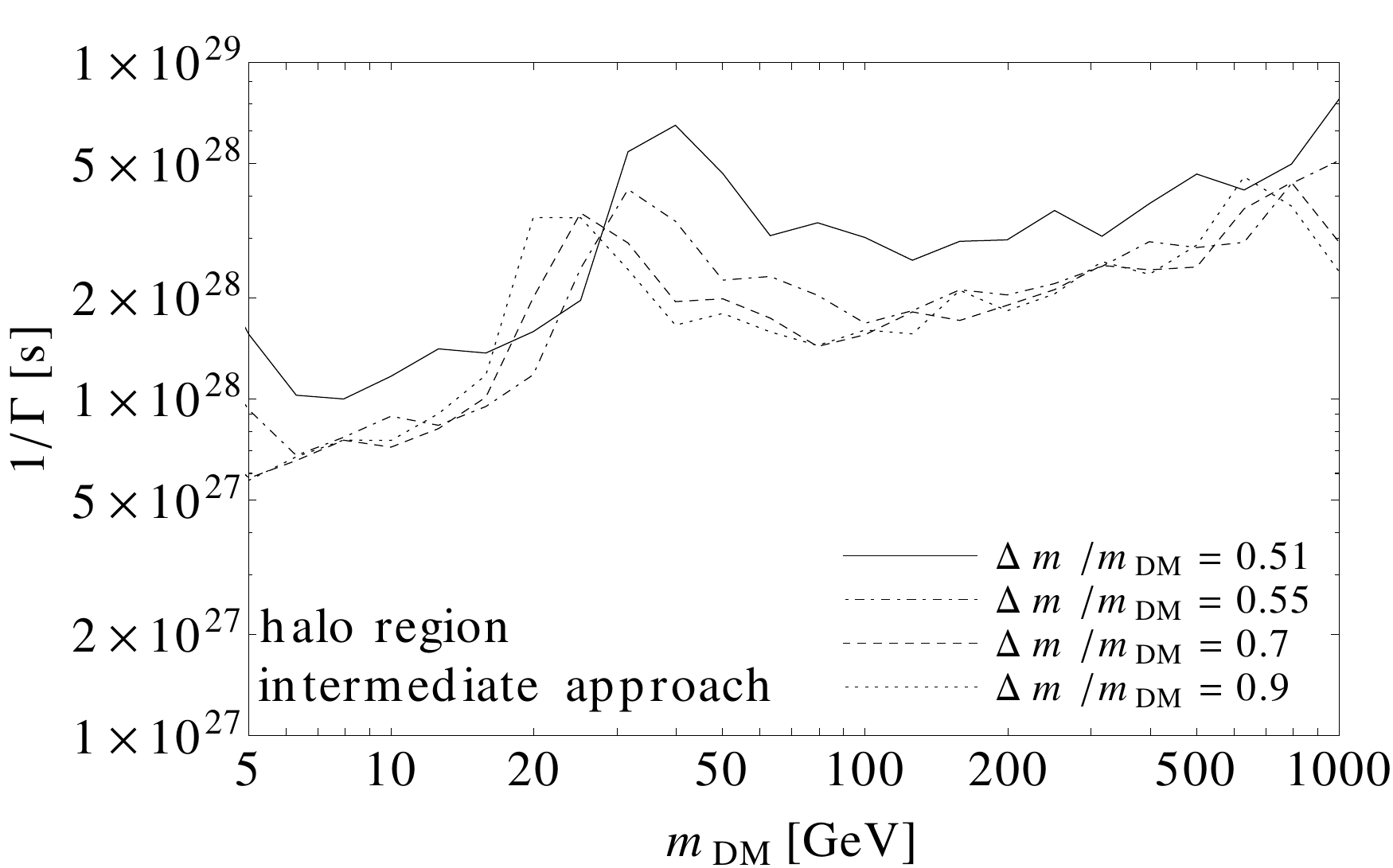}
\caption{
Gamma-ray constraints on cascade dark matter decays for the halo region. The lines show our constraints in the $\Gamma^{-1}$ vs $m_{DM}$ plane for several values of the degeneracy parameter $\Delta m/m_{DM}$ and using the intermediate approach.
}
\label{fig:constraints_dec}
\end{figure}

 Figure \ref{fig:constraints_dec} displays our intermediate constraints on the decay timescale $\Gamma^{-1}$ for decaying dark matter taking Fermi-LAT observations of the halo region. As in the case of annihilations, it is clear that cascade decays into photons place relatively strong limits on dark matter models. In particular, the lower limits on $\Gamma^{-1}$ fall in the range $10^{28}-10^{29}$ s , i.e.~two to three orders of magnitude above the typical $10^{26}$ s used in the literature \cite{Ibarra:2008jk} to meet the PAMELA positron excess. The bumps at $m_{DM}=20-40$ GeV present in figure \ref{fig:constraints_dec} have the same origin as the dips in figure \ref{fig:constraints_ann} (left) which have been explained above.

\par Up to this point we have been focussing on the constraints provided by the search for box-shaped gamma-ray features in Fermi-LAT data. Let us now briefly comment on the detectability of these type of signatures. In the case of the widely-studied gamma-ray lines, a positive, statistically significant detection would give the peak energy and amplitude of the line which would translate directly into $m_{DM}$ and $\langle \sigma v \rangle$ (modulo background modelling and uncertainties regarding the $J-$factor). In the case of a box spectrum, instead, one would probably just spot the high-energy shoulder and accordingly estimate the box edge $E_{+}$ and amplitude. These two observables are obviously degenerate in the parameter space $(m_{DM},\langle \sigma v \rangle,\Delta m/m_{DM})$. Thus, the discovery of a hard shoulder would be somewhat problematic in reconstructing the corresponding dark matter phenomenological parameters and in connecting to concrete particle physics models. Nevertheless, it rests as a possibility that such spectral feature is detected at very high energies.

\section{Concrete models}\label{secModels}

We consider a model where the dark matter is constituted by complex scalar particles, $\chi$, which are singlets under the Standard Model gauge group. To ensure the stability of the dark matter particle we introduce a discrete $Z_2$ symmetry, under which $\chi$ is odd while all the Standard Model particles are even. The Lagrangian then reads  ${\cal L}={\cal L}_{\rm SM}+{\cal L}_{\chi}+{\cal L}_{\rm int}$, where ${\cal L}_{\chi}$ contains the kinetic term and the potential term of the extra scalar $\chi$ and ${\cal L}_{\rm int}$ is the interaction term of $\chi$ and the Standard Model particles, which simply reads:
\begin{equation}
-{\cal L}_{\rm int}=\lambda |\chi|^2 |H|^2\quad,
\end{equation}
where $H$ is the Standard Model Higgs doublet, $H=\frac{1}{\sqrt{2}}(0,v+h)$, with $v=246$ GeV. In this simple model, the dark matter particle can annihilate $\chi\chi\rightarrow h h$ with a sizable cross-section~\cite{McDonald:1993ex}. Subsequently, the Higgs particle $h$ can decay $h\rightarrow \gamma\gamma$, with a branching ratio which is $\sim 2\times 10^{-3}$ for the mass window of the light Higgs $m_h\simeq 110-130$ GeV \cite{Djouadi:2005gi}, presently allowed by searches at the Large Hadron Collider \cite{atlasH,cmsH}. In this case, the branching ratio $h\rightarrow \gamma\gamma$ is too small to allow the observation of a gamma-ray signal, if the annihilation cross-section coincides with the thermal cross-section $\langle \sigma v\rangle \simeq 3\times 10^{-26}\,{\rm cm}^3/{\rm s}$. Nevertheless, it follows from figure \ref{fig:constraints_ann} that a signal could be observed if the boost factor for $\chi\chi \to h h$ is ${\cal O}(20)$. Boost factors of this order are not unfeasible if a combination of substructure \cite{Lavalle:1900wn}, Sommerfeld enhancement \cite{Hisano:2003ec} and/or non-standard cosmological scenarios \cite{Schelke:2006eg} is at play. Furthermore, this limit on the boost factor is similar to the limit obtained in \cite{Papucci:2009gd} for annihilations into Higgses from the first year of Fermi-LAT gamma-ray observations in the full sky, although it is likely weaker when compared to dwarf \cite{Ackermann:2011wa} and halo \cite{Ackermann:2012rg} recent observations. This simple example shows the power of the search for box-shaped gamma-ray features in constraining dark matter models. 

The intensity of the gamma-ray box signal can be enhanced by considering the following extension of the previous model. In addition to the dark matter particle $\chi$, we introduce a second, real, scalar field $\phi$, singlet under the Standard Model gauge group and two fermions $\xi_1$ and $\xi_2$, singlets under $SU(3)_c\times SU(2)_L$ and with hypercharge $Y=1$ and $-1$ respectively (the model is then free of gauge anomalies). Furthermore, we assume that $\phi$ is even under the discrete $Z_2$ symmetry, while $\xi_1$ and $\xi_2$ are odd.\footnote{Assigning an odd charge to $\xi_i$ under $Z_2$ is not determined by the dark matter physics, but to prevent the mixing of the right-handed electron with $\xi_2$.} The Lagrangian then reads  ${\cal L}={\cal L}_{\rm SM}+{\cal L}_{\chi}+{\cal L}_\phi+ {\cal L}_\xi+{\cal L}_{\rm int}$. Here, ${\cal L}_{\chi}$ and ${\cal L}_{\phi}$ are the Lagrangians for a complex and real scalar fields, respectively, ${\cal L}_\xi$ contains the kinetic terms for the fermion fields $\xi_1$ and $\xi_2$, and ${\cal L}_{\rm int}$ contains the interaction terms among fields:
\begin{equation}
\begin{split}
-{\cal L}_{\rm int}&=  y \bar \xi_1 \xi_2 \phi+ m_\xi \bar\xi_1\xi_2+
e \bar \xi_1 (A_\mu - \tan\theta_W Z_\mu) \gamma^\mu\xi_1
-e \bar \xi_2(A_\mu - \tan\theta_W Z_\mu)\gamma^\mu \xi_2+{\rm h.c.} \nonumber \\
&+\frac{1}{4}\lambda_{\phi H} \phi^2|H|^2 +\frac{1}{4}\lambda_{\chi H} |\chi|^2|H|^2 +\frac{1}{4}\lambda_{\phi \chi} \phi^2|\chi|^2+ \mu \phi |H|^2 + \alpha \phi |\chi|^2  \quad.
\end{split}
\label{L_int}
\end{equation}

With this interaction Lagrangian, the dark matter particle $\chi$ can annihilate into two scalar particles, $\chi\chi\rightarrow \phi\phi$, as well as into $h h$, $ZZ$, $WW$ or $f\bar f$~\cite{McDonald:1993ex}. Assuming that the coupling $\lambda_{\phi\chi}$ is much larger than $\lambda_{\chi H}$, then the dark matter particle predominantly annihilates into the intermediate $\phi$-scalars. 

The annihilation is followed by the decay of $\phi$. At tree level $\phi$  can only decay into two Higgs particles $\phi\rightarrow h h$, provided the decay is kinematically allowed, namely when $m_\phi>2m_h$. In this case the decay rate is
\begin{equation}
\Gamma(\phi\rightarrow h h)=\frac{m_\phi \mu^2}{128\pi}
\left(1-\frac{4m_h^2}{m_\phi^2}\right)^{1/2}\quad.
\end{equation}
This is usually the dominant decay mode, unless the coupling $\mu$ is very small.

At the one loop level, on the other hand, new decay channels are possible and can have a sizable branching ratio when the decay $\phi\rightarrow h h$ is kinematically forbidden. Since $\xi_{1,2}$ carry hypercharge, $\phi$ can decay via an one loop diagram into $\gamma\gamma$ and, if kinematically allowed, also into $\gamma Z$ and $ZZ$. The rates for the decay processes into gauge bosons read, in the limit $m_\xi\gg M_Z,~m_\phi$,
\begin{equation}
\begin{split}
\Gamma(\phi\rightarrow\gamma\gamma)&\simeq
\frac{y^2 \alpha^2}{144\pi^3}
\frac{m_\phi^3}{m_\xi^2}\quad,\\
\Gamma(\phi\rightarrow\gamma Z)&\simeq
\frac{y^2 \alpha^2 \tan^2\theta_W}{72\pi^3}
\frac{m_\phi^3}{m_\xi^2}
\Big(1-\frac{M^2_Z}{m_\phi^2}\Big)^3\quad, \\
\Gamma(\phi\rightarrow Z Z)&\simeq\frac{y^2 \alpha^2 \tan^4\theta_W}{144\pi^3}\frac{m_\phi^3}{m_\xi^2}
\left(1-\frac{4 M_Z^2}{m_\phi^2}\right)^{1/2} \left[\left(1-\frac{2M_Z^2}{m_\phi^2}\right)^2
+\frac{1}{2}\left(\frac{2M_Z^2}{m_\phi^2}\right)^2\right]\quad.
\end{split}
\end{equation}

Furthermore, the scalar $\phi$ can decay $\phi\rightarrow b\bar b$ via a loop diagram involving the coupling $\frac{1}{2}\mu \phi h h$ to the Higgs particle. The decay rate for this process reads
\begin{equation}
\Gamma(\phi\rightarrow b\bar b)=\frac{m_\phi}{128\pi^3}
\left(\frac{G_F m_b^3 \mu}{m_h^2}\right)^2\left|I\left(\frac{m_\phi^2}{4m_h^2}\right)\right|^2\quad,
\end{equation}
where, taking $m_b\ll m_h$,
\begin{equation}
I(w)=\int_0^1 dx\,\int_0^{1-x}dy\,\frac{1-x-y}{x+y-4 w x y}\quad.
\end{equation}

\begin{figure}[htp]
\centering
\hspace{-1cm}
\includegraphics[width=0.52\textwidth,height=0.43\textwidth]{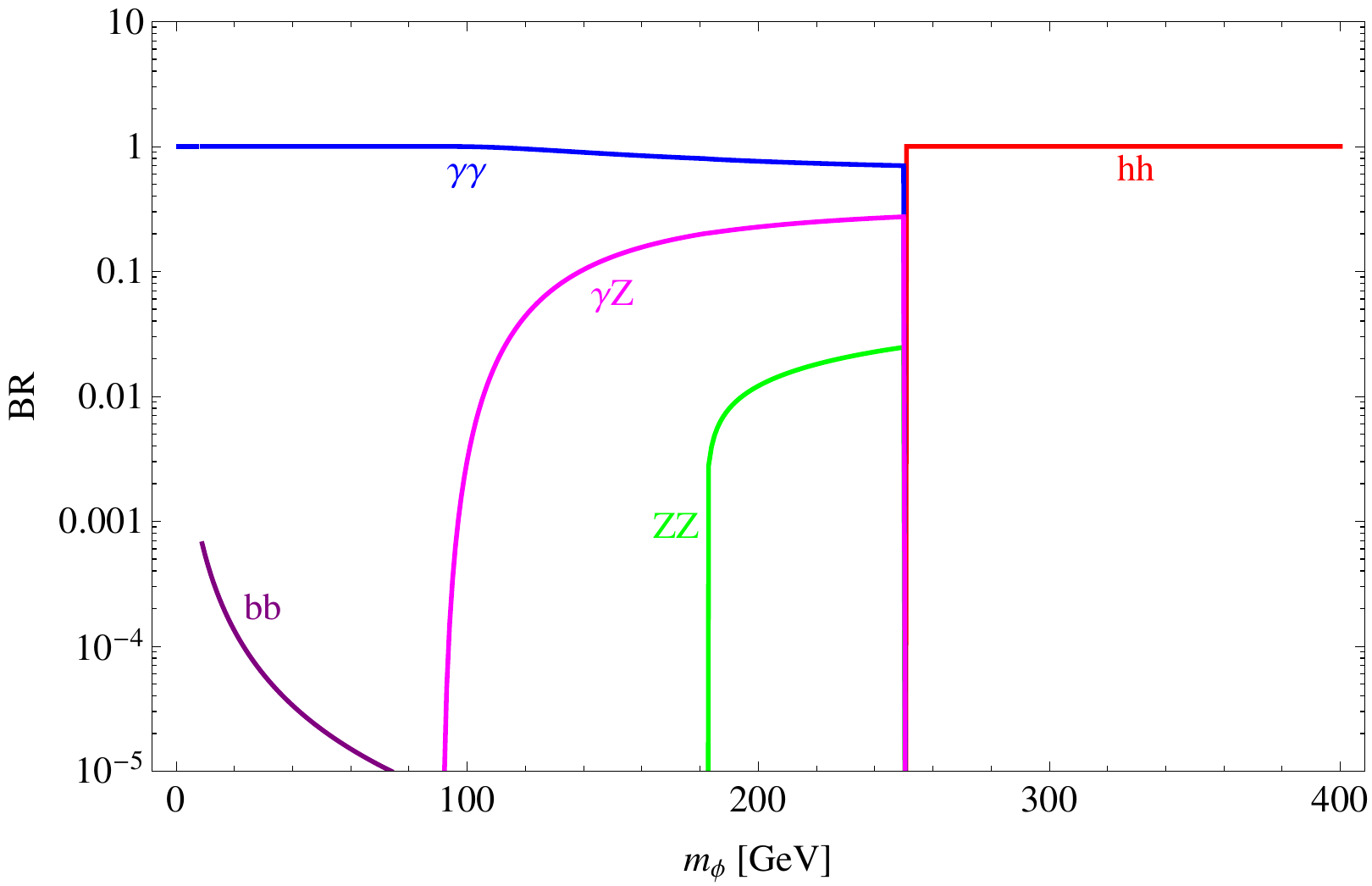}
\caption{Branching ratios for the decay of the intermediate scalar particle $\phi$ for the concrete model described in the text. 
}
\label{fig:BRs}
\end{figure}

We show in figure \ref{fig:BRs} the branching ratios for $\phi$ decays in this model, assuming for concreteness $y=1$, $m_\xi=500 $ GeV, $\mu=100$ GeV and $m_h=125$ GeV. As apparent from the figure, when $m_\phi< 2m_h$, the scalar $\phi$ decays dominantly into two photons. Therefore, in this situation the stringent constraints in the parameter space presented in figure \ref{fig:contour} (left) apply. Let us note however that in this particular model the dark matter particle is not Majorana, and accordingly the flux in equation \eqref{unconvol} is to be multiplied by 1/2.

The study of more complex particle physics models in which box constraints are relevant is deferred to future work.

\section{Conclusion}\label{secConcl}
\par In this work we have paid close attention to a class of gamma-ray spectral signatures that, to the best of our knowledge, has not been explored in the framework of dark matter searches. Cascade dark matter annihilations or decays into photons give rise to a box-shaped gamma-ray spectrum that resembles a line in the limit of degenerate dark matter and intermediate scalar masses, and a flat plateau in the non-degenerate case. This is the feature of interest for the present paper. The key point is that, as the mass splitting increases, the photon spectrum gets dimmer but reaches higher energies, yielding sizable fluxes even for highly non-degenerate configurations. We have illustrated how this unique phenomenology applies to dark matter searches by using the 2-yr Fermi-LAT data. Overall, our constraints for both annihilating and decaying dark matter are rather strong, reaching $\langle \sigma v \rangle\sim10^{-29} \textrm{ cm}^3\textrm{/s}$ (i.e.~$\sim3\times10^{-4}$ in branching ratio assuming a thermal cross-section) and $\Gamma^{-1}\sim 3\times10^{29}$ s in the most aggressive cases. Also, we have found a rapid ``saturation'' of the constraints as the mass splitting between dark matter and intermediate scalar is increased. Concretely, any mass splitting $\Delta m/m_{DM}$ above $\sim$5\% leads essentially to the same annihilation cross-section upper limits, which are only a few times weaker than the degenerate configuration. A similar situation holds for the decaying dark matter constraints. This means that no fine-tuning between the masses at play is required to place stringent limits on the properties of dark matter. Consequently, the non-observation of box-shaped gamma-ray features is extremely effective in constraining concrete particle physics frameworks. We point out that more detailed research regarding this alternative spectral signature is needed in view of the exquisite data provided by the Fermi-LAT instrument as well as the higher energy data from IACTs and, eventually, CTA.

%\vspace{0.5cm}
{\it Acknowledgements:} 
The authors would like to thank Christoph Weniger for fruitful discussions. This work was partially supported by the DFG cluster of excellence ``Origin and Structure of the Universe''.

%%%%%%%%%%%%%%%%%

\bibliographystyle{apsrev.bst}
\bibliography{gammabox}

\end{document}